\documentclass[doublecol]{epl2}
\usepackage{amssymb}
\usepackage{amsmath}

\title{Ultrasonic studies of the magnetic phase transition in MnSi}
\shorttitle{Title} 

\author{A.E. Petrova and S. M. Stishov \thanks{E-mail: \email{sergei@hppi.troitsk.ru}}}
\shortauthor{A. E. Petrova \etal}

\institute{
    \ Institute for High Pressure Physics - Troitsk, Moscow
Region, Russia\\
 }

\abstract{Measurements of the sound velocities in a single crystal of MnSi were performed in the temperature range 4-150 K. Elastic constants, controlling propagation of longitudinal waves reveal significant softening at a temperature of about 29.6 K and small discontinuities at $\sim$28.8 K, which corresponds to the magnetic phase transition in MnSi. In contrast the shear elastic moduli do not show any softening at all, reacting only to the small volume deformation caused by the magneto-volume effect. The current ultrasonic study exposes an important fact that the magnetic phase transition in MnSi, occurring at 28.8 K, is just a minor feature of the global transformation marked by the rounded maxima or minima of heat capacity, thermal expansion coefficient, sound velocities and absorption, and the temperature derivative of resistivity.}

\begin{document}

\maketitle
The nature of the helical magnetic phase transition in MnSi, occurring at about 29 K at ambient pressure, has been a subject of significant controversy. For a long time the phase transition in MnSi at ambient pressure was generally believed to be second order or continuous, though there have been no proper proofs for that conclusion. In this context, the evolution of magnetic susceptibility of MnSi at high pressure~\cite{1} also seemingly indicated a tricritical point on the phase transition line, where the second order phase transition became first order. As a result, the idea was borne that tricritical behavior was a generic feature of itinerant ferromagnets at low temperatures~\cite{2}. Recent precise measurements of the lattice constants of MnSi apparently confirmed a first order character of the phase transition in MnSi at high pressure and low temperature~\cite{3}. However, studies of thermodynamic and transport properties of a high quality single crystal of MnSi~\cite{4,5} strongly suggested a first order nature of the magnetic phase transition in MnSi even at ambient pressure, a result in direct contrast with the previous assumption.

But arguments presented in~\cite{4,5} were not quite complete because in real solids inevitable imperfections will almost always smear first order phase transitions over certain temperature and pressure intervals. Consequently, variations of heat capacity and thermal expansion in the best case would assume the form of a broadened $\delta $--function that might be interpreted as evidence for divergence typical of second order phase transitions.

 In Refs.~\cite{4,5}, sharp peaks in heat capacity, thermal expansion coefficient and temperature derivative of resistivity in a temperature interval less than 0.1 K signified the magnetic phase transition in MnSi. Sharpness of the peaks and obvious lack of mean-field-like anomalies favor a first order nature of the phase transition. To confirm this point, it is desirable to perform an experiment that minimizes effects of smearing the phase transition. Ultrasonic studies of the phase transition in MnSi seems most appropriate in this sense, because sound velocities, which are relevant to this case, are defined by local values of elastic constants and their evaluation does not involve a procedure that requires probing certain ranges of temperature or pressure. The high sensitivity and potential accuracy of ultrasonic measurements make them a valuable tool for studying subtleties of phase transitions.

There are three possible scenarios for the behaviors of sound velocities and sound attenuation at a phase transition~\cite{6,7,8}:

\noindent a) Finite jumps in sound velocity and attenuation coefficient are expected at a second order, mean-field phase transition.

\noindent b) A power-law divergence in sound velocity and attenuation coefficient are characteristic of a second order phase transition with strong fluctuations. The corresponding critical exponent for sound velocity is expected to be equal that of heat capacity.

\noindent c) At a first order phase transition, there will be a finite jump of sound velocity and $\delta $--function like behavior of the attenuation coefficient.

\begin{figure}[htb]
\includegraphics[width=86mm]{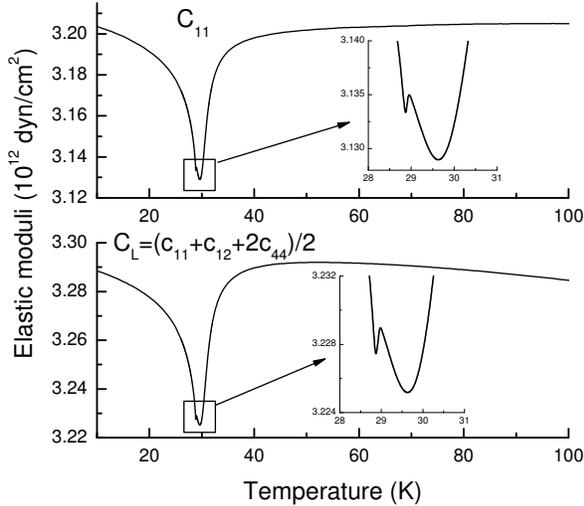}
\caption{\label{fig1}Elastic moduli of MnSi, controlling propagation of longitude sound waves.}
\end{figure}
\begin{figure}[htb]
\includegraphics[width=86mm]{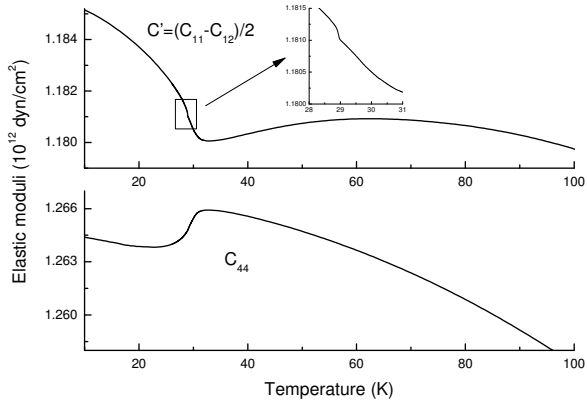}
\caption{\label{fig2} Shear elastic moduli of MnSi.}
\end{figure}

 We report results of ultrasonic studies of a single crystal of MnSi in the temperature range 4-150K. In the course of these studies, several runs were performed using digital pulse-echo techniques (see~\cite{9} and references therein). Three samples of MnSi of 4.7, 2.15 and 5.05 mm thicknesses and with orientations along [110] and [100] were cut from a high quality big single crystal, characterized previously~\cite{4,5}. The corresponding surfaces of the samples were made optically flat and parallel. The 36$^\circ$ Y ( P-wave) and 41$^\circ$ X (S-wave) cut $LiNbO_3$ transducers were bonded to the samples with silicon grease. Temperature was measured by a calibrated Cernox sensor with an accuracy of 0.02 K.

 A sinusoidal pulse of $\sim$50 MHz was sent to the transducer that excites a sound wave, experienced multiple reflections inside the sample. The distance between two arbitrary reflections corresponds to a multiple of the round trip travel time, which is determined by performing a cross-correlation between two selected reflections. The speed of sound and elastic constants are calculated using the known thickness and density of the samples and the relationship $c_{ij} = \rho V^2$. The precision of the sound velocity determinations is no worse than one part in $10^6$; whereas, the absolute accuracy is about 0.1\% due to uncertainty connected with a phase shift at the transducer--sample bond. The precision and accuracy of the calculated elastic constants are basically comparable but may be not as good when calculated using data from different runs with the samples of different orientations.

\begin{figure}[htb]
\includegraphics[width=86mm]{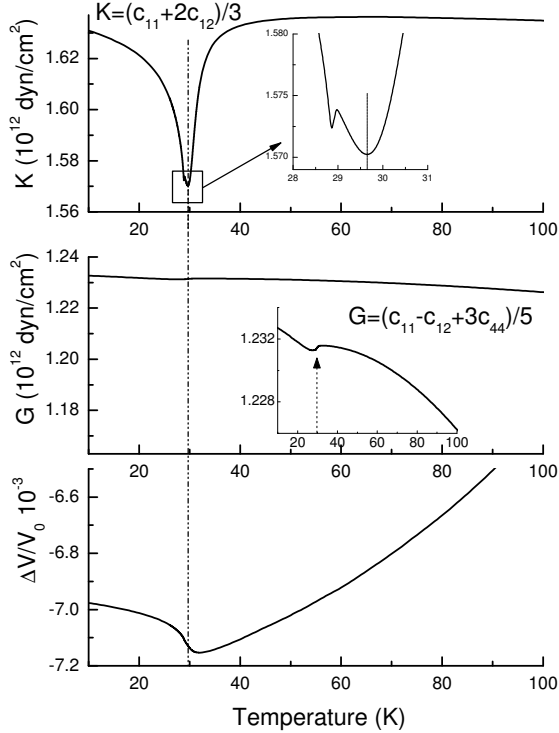}
\caption{\label{fig3} Bulk modulus (K), average shear modulus (G) and volume change ($\Delta V/V_o$) of MnSi~\cite{5}. Averaging is done according to the Voigt approximation~\cite{13}. The dash--dot line and arrow indicate position of the rounded minimum of K.}
\end{figure}

\begin{figure}[htb]
\includegraphics[width=86mm]{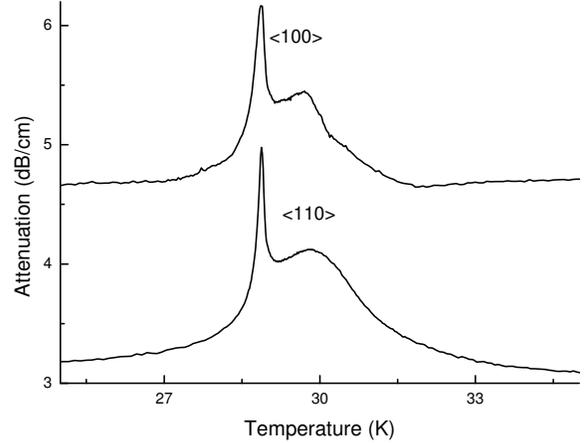}
\caption{\label{fig4}Attenuation of the longitudinal sound waves in MnSi. No background is subtracted.}
\end{figure}

The elastic constants c$_{11}$ and $c_{44}$ and the combinations $C_L= (c_{11}+c_{12}+2c_{44})/2$ and $C^\prime = (c_{11}-c_{12})/2$ were obtained directly from data on longitudinal and transverse sound velocities along [100] and [110] directions (fig.~\ref{fig1},~\ref{fig2}, see also~fig.~\ref{fig3}). The sound attenuation is estimated from an analysis of amplitudes of two adjacent reflections as $20/L\log_{10}(A1/A2)$ (fig.~\ref{fig4}). The precision of the attenuation data is about 0.5 \%, though the absolute accuracy is uncertain. These data are qualitatively comparable with low resolution results on sound velocity~\cite{10} and sound attenuation measurements~\cite{11}. It is of interest to compare the present data on MnSi with elastic properties of its structural analog FeSi~\cite{12}; numerical values of the corresponding elastic constants appear to be rather close in both materials.

As is shown in fig.~\ref{fig1}, the elastic constants c$_{11}$ and $C_L= (c_{11}+c_{12}+2c_{44})/2$, controlling propagation of longitudinal waves in the [100] and [110] directions, reveal pronounced rounded dips at about 29.6 K, with discontinuities at the low temperature side of the dips at $\sim$28.8 K, which corresponds to the magnetic phase transition in MnSi. Both of these features are beautifully correlate with those in heat capacity, thermal expansion and resistivity data obtained with the same single crystal of MnSi in~\cite{4}, except sharp peaks in these thermodynamic and transport properties are replaced by a modest discontinuity in elastic properties at the phase transition. In contrast, the shear elastic moduli $c_{44}$  and $C^\prime = (c_{11}-c_{12})/2$ do not exhibit any dips at all, reacting only to a small volume deformation caused by the magneto-volume effect (see fig.~\ref{fig2}). Note that the tiny volume change at the magnetic phase transition in MnSi~\cite{4} is reflected in behavior of $C^\prime = (c_{11}-c_{12})/2$ (see inset in fig.~\ref{fig2}).

Generalized features of the elastic properties of MnSi are illustrated in fig.~\ref{fig3}, which displays behavior of the bulk modulus $K=(c_{11}+2c_{12})/2$, the average shear modulus $G=(c_{11}-c_{12}+3c_{44})/5$~\cite{13} and the volume change $\Delta V/V_o$. It is clearly seen in fig.~\ref{fig3} that the small drop in G is associated with the volume increase resulting from magnetic ordering and does not carry any traces of softening as the bulk modulus certainly does. Moreover, the total variation of the shear modulus G is an order of magnitude less than the variation of the bulk modulus K in the phase transition region. Bearing in mind the itinerant nature of magnetism in MnSi, this drastic difference in behavior of the longitudinal and transverse elastic properties may imply the essential decoupling of electron and ionic subsystems in this material in respect to the magnetic transformation.

 Now we turn to the sound attenuation. Two curves illustrating attenuation of longitude waves in [100] and [110] directions \footnote{Our data on attenuation of transverse waves indicate the existence of small but distinct maxima at a temperature of about 29~K.} are shown in fig.~\ref{fig4}. The double peak structure makes these curves look like almost exact copies of those characterizing behavior of the heat capacity, thermal expansion and resistivity in the vicinity of the phase transition in MnSi~\cite{4,5}. It needs to be pointed out that the sharp peaks in sound attenuation are quite symmetric and do not look like they should to be associated with a second order phase transition~\cite{6,7,8}. This behavior of the attenuation most probably is connected with violation of adiabatic conditions arising from a finite entropy change at the first order phase transition in MnSi. In this case, the sound attenuation should behave similar to a heat capacity curve at a slightly rounded first order phase transition that can be described as a smeared $\delta $--function~\cite{4}. To this end, all features observed match to the case c), corresponding to a first order phase transition.

At the same time, the current ultrasonic study exposes another important fact, making obvious that the magnetic phase transition in MnSi, occurring at 28.8 K, is just a minor feature of the global transformation that is marked by the rounded maxima or minima of heat capacity, thermal expansion coefficient, sound velocities and absorption, and the temperature derivative of resistivity~\cite{4,5}. Behavior of the bulk modulus of MnSi (fig.~\ref{fig3}) indicates a tendency toward a volume instability that may characterize this transformation as an incomplete second order transition. But, the nature of the global transformation still remains a puzzle, though a number of plausible spin structures (blue quantum fog~\cite{15}, spin crystal~\cite{16} and skyrmion textures~\cite{17}), suggested for the paramagnetic phase of MnSi may be pertinent. On the other hand, relevant experimental data are too scarce to solve the problem unambiguously. What we actually know from neutron scattering experiments~\cite{18,19,20} is that, on cooling MnSi and its closest analog FeGe, a quasi-long-range helical order (QLRO) forms at temperatures slightly above the phase transition temperature. However, for yet unknown reasons the state with QLRO can not reach a true long range order in the continuous way. This may suggest identification of the QLRO state as a frustrated magnetic state, possibly emerging as a result of competition between different directions of the spin helix~\cite{5,20}. Consequently, a first order phase transition is needed to unlock the situation.

\acknowledgments
In conclusion, we would like to acknowledge valuable advice of  Cristian Pantea, Steven Jacobsen,  Baosheng Li and Yuri Pisarevskii concerning the ultrasonic equipment. Elena Gromnitskaya, Cristian Pantea and Izabela Stroe made some preliminary measurements of the elastic properties of MnSi. Technical help of Vladimir Krasnorusski is greatly appreciated. Thomas Lograsso provided the single crystal of MnSi. Authors are thankful  to  J.D. Thompson for valuable remarks. We appreciate support of the Russian Foundation for Basic Research (grant 09-02-00336), Program
of the Physics Department of RAS on Strongly Correlated Systems and Program of the Presidium of RAS on Physics of Strongly
Compressed Matter.

\end{document}